\documentclass[12pt,a4paper,nofootinbib]{revtex4}

\usepackage{graphicx}

\usepackage[fleqn,sumlimits,intlimits]{amsmath}
\usepackage{latexsym,amssymb}
\usepackage[english]{babel}
\usepackage{verbatim}
\usepackage{color}

\newcommand{\bea}{\begin{eqnarray}}
\newcommand{\eea}{\end{eqnarray}}
\newcommand{\beq}{\begin{equation}}
\newcommand{\eeq}{\end{equation}}
\newcommand{\nn}{\nonumber}
\newcommand{\bal}{\begin{align}}
\newcommand{\eal}{\end{align}}
\newcommand{\bit}{\begin{itemize}}
\newcommand{\eit}{\end{itemize}}
\newcommand{\rar}{\rightarrow}

\newcommand{\abs}[1]{\vert #1\vert}
\newcommand{\avg}[1]{\left\langle #1 \right\rangle}

\newcommand{\half}{{\frac{1}{2}}}

\newcommand{\dS}{{d_s}}

\newcommand{\leaveout}[1]{}

\newcommand {\Tfour}{{ {\cal T}_4 }}

\newcommand{\sg}{\sigma}
\newcommand{\prt}{\partial}
\newcommand{\cD}{{\cal D}}

\newcommand{\meas}{{\mu(\nu)}}

\usepackage{amsthm,amsmath,amssymb,mathrsfs}

\begin{document}

\title{Dynamical dimensional reduction in toy models of 4D causal quantum gravity}

\author{Georgios Giasemidis$^{a}$}
\author{John F Wheater$^{a}$}
\author{Stefan Zohren$^{a,b}$}

\affiliation{
$^{a}$Rudolf Peierls Centre for Theoretical Physics, 1 Keble Road, Oxford OX1 3NP, UK \\
$^{b}$Department of Physics, Pontifical Catholic University of Rio de Janeiro, Rua Marqu\^es de S\~ao Vincente 225, Rio de Janeiro, Brazil}

\date{Revised August 28, 2012}


\begin{abstract}
In recent years several approaches to quantum gravity have found evidence for a scale dependent spectral dimension of space-time varying from four at large scales to two at small scales of order of the Planck length. The first evidence came from numerical results on four-dimensional causal dynamical triangulations (CDT) [Ambj{\o}rn et al.,\ Phys. Rev. Lett. \textbf{95} (2005) 171]. Since then little progress has been made in analytically understanding the numerical results coming from the CDT approach and showing that they remain valid when taking the continuum limit. Here we 
argue that the spectral dimension can be determined from a model with fewer degrees of freedom obtained from the CDTs by ``radial reduction". In the resulting ``toy" model we can take the continuum limit analytically and obtain a scale dependent spectral dimension varying from four to two with scale and having functional behaviour exactly of the form which was conjectured on the basis of the numerical results.
\end{abstract}

\maketitle

\section{
Introduction}
The quest to reconcile classical general relativity and quantum mechanics has a long history. One of the main difficulties is the fact that gravity is perturbatively non-renormalizable in four dimensions as was shown by 't Hooft and Veltman in the seventies \cite{hooft}. However, defining quantum gravity non-perturbatively, there is evidence from different approaches that there might still be a non-trivial ultraviolet fixed-point as suggested by Weinberg \cite{weinberg}. The causal dynamical triangulation (CDT) approach to quantum gravity (see \cite{Ambjorn:1998xu} and \cite{review} for a review) recently gave a surprising answer to what dynamical mechanism might regulate the theory at short distances in such a scenario. In particular, numerical simulations \cite{Ambjorn:2005db} show evidence for the dynamical reduction of the spectral dimension from four at large scales to two at small scales of order of the Planck length (see also \cite{dario} for a discussion in a three-dimensional setting). More recently similar results have also been observed in other approaches to quantum gravity, most notably the exact renormalization group approach \cite{LL} and Ho\v{r}ava-Lifshitz gravity \cite{Horava}. This also points towards several possible relations between those approaches \cite{HLrel}. 

\section{
Previous numerical insights from four-dimensional CDT
}
The idea behind the spectral dimension as a scale dependent measure of dimensionality is the following. Consider a diffusion process on a fixed space-time geometry. The diffusion kernel $K_g$ is determined by the diffusion equation
\bea\label{ja2}
\frac{\prt}{\prt \sg} \, K_g(y,y_0;\sg) = \Delta_g K_g (y,y_0;\sg),
\eea
 where $g$ denotes the space-time metric, $y_0$ the starting point of the diffusion and $y$ the position of the diffusion process after time $\sigma$. We can then define the return probability (density) by choosing starting and end point equal and integrating over all positions
\bea
P_{g}(\sg) = \frac{1}{V_g} \int d^dy \sqrt{\det g_{ab}(y)} \; K_g(y,y;\sg)
\eea
with $V_g=\int d^dy \sqrt{\det g_{ab}(y)}$. If we consider diffusion on quantum space-time we would have to take the ensemble average, formally defined using the gravitational path integral
\bea \label{expecP}
\avg{P(\sigma)}_Z =
\frac{1}{Z} \int  \cD [g_{ab}] \; e^{-{S}_E(g_{ab}) }  \, P_g(\sg),
\eea
where $Z = \int  \cD [g_{ab}] \; e^{-{S}_E(g_{ab}) }$ is the partition function and ${S}_E(g_{ab})$ the Euclidean Einstein Hilbert action.
Formally, the scale dependent spectral dimension is then defined as \cite{Ambjorn:2005db}
\bea
D_s(\sigma) =-2\frac{d \log \avg{P(\sigma)}_Z}{d\log\sigma},
\eea
where $\sigma$ is the diffusion time and corresponds to the scale at which the diffusion process probes the quantum geometry. 

\begin{figure}
\begin{center}
\includegraphics[width=4in]{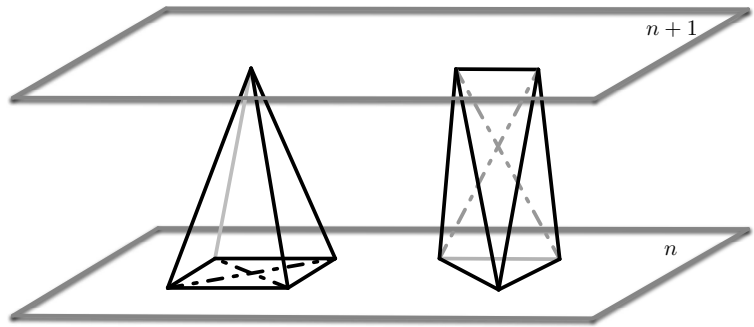}
\caption{The two building blocks of four-dimensional CDT; The (4,1) and (3,2) simplex are on the left and right respectively. Dotted lines correspond to 3 simplices.}
\label{fig1}
\end{center}
\end{figure}

The set of four-dimensional rooted causal triangulations  $\Tfour$ consists of all graphs $T$ of topology $I\times \Sigma^3$ made of (4,1) and (3,2)-simplices connecting vertices at distance $n$ to vertices at distance $n+1$ from the root, see Figure \ref{fig1} \cite{cdtconst}.  $T$  is partly characterised by the number of $i$-simplices,  $N_i(T)$, it contains; as well as two different four-simplices, there are three different types of three-simplices and two different types of triangles and links (i.e.\ space-like and time-like). These ten quantities are related by seven topological constraints so only three are independent \cite{cdtconst}. In the CDT approach the gravitational partition function is defined as
\bea  \label{partition}
Z=\sum_{T\in\Tfour} \frac{1}{C_T} {\rm e}^{-S_E(T)},
\eea
where $C_T$ is a combinatorial symmetry factor and 
the Euclidean Einstein-Regge action of the triangulation $T$ is
\bea
S_E(T) = \lambda N_4(T) -\nu N_2(T).
\eea
Here $\lambda$ is the bare cosmological constant and $\nu$ is the bare inverse Newton's constant. 

To compute the spectral dimension one should evaluate \eqref{expecP} using the partition function \eqref{partition} and
then take the infinite volume limit in which $\lambda$ is tuned towards its  critical value and $\nu$ is expressed in terms of the  inverse renormalized Newton's constant $1/G$. At present this is analytically out of reach but Monte Carlo simulations of random walks (discrete diffusion) on four-dimensional CDTs of fixed $N_4$ \cite{Ambjorn:2005db} yield a scale dependent spectral dimension given by
\bea
D_s(\sigma)=  4.02-\frac{119}{54+\sigma}=
\begin{cases}
1.80 \pm 0.25, & \sigma\to 0,\\
4.02 \pm 0.1, & \sigma \to \infty \label{DsJan}.
\end{cases} 
\eea

A possible objection to \eqref{DsJan} is that the simulations are inevitably affected by finite size effects and the dimensional reduction observed might simply be an artefact of the discreteness scale. However, assuming that this expression can be extrapolated to continuum physics,  the return probability (density) \eqref{expecP} for four-dimensional CDT in the \emph{infinite volume limit} \cite{Ambjorn:2005db} would be
\bea  \label{intro-P}
\avg{P(\sigma)}_Z \sim \frac{1}{\sigma^2}  \frac{1}{ 1 + const.\, G / \sigma},
\eea
where ``$\sim$" denotes equality up to multiplicative logarithmic corrections.

\section{Random walk on CDT}

\begin{figure}
\begin{center}
\includegraphics[width=4in]{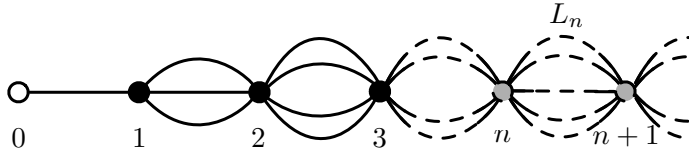}
\caption{An example of a multigraph.}
\label{fig2}
\end{center}
\end{figure}

The strict sliced structure of CDT graphs implies that every step in a random walk either moves in the time-like direction or to a neighbouring vertex in the same spatial hypersurface. We can then consider a projection of the random walk in which it is viewed as one-dimensional (in the  time direction) with biases caused by the connectivity between adjacent hypersurfaces  and a delay in the diffusion time caused by excursions into the spatial hypersurface.  The delays will generally be different for each visit to a given hypersurface but if 
 the number of space-like edges attached to any vertex is bounded by a  finite number then the delay time will be finite and the detailed structure of the spatial hypersurfaces cannot affect the spectral dimension  -- it can only produce a  finite rescaling of the time variable. For the 2d CDT,  where there are exactly two space-like edges attached to each vertex, this intuition has been turned into a rigorous  argument \cite{Durhuus:2009sm}. 
 The spectral dimension of a 2d CDT graph $T$ is bounded above by that of the multigraph $M$ obtained from radial reduction of the full graph  
which maps  all vertices in $T$ at a fixed graph distance $n$ from the root to a single vertex while retaining all time-like edges (see Figure \ref{fig2}). Denoting the number of time-like edges connecting vertex $n$ to vertex $n+1$ by $L_n$,  $M$ is characterised by the sequence $M=\{L_n,n=0,1,2, ...\}$. The partition function \eqref{partition} induces a measure $\meas$ for 
$M$ which defines the multigraph ensemble. 

 There are  two relevant  characteristics which determine the spectral dimension of $M$; the growth in the number of time-like edges with distance from the root and the behaviour  of the electrical resistance of the graph viewed as an electrical network  \cite{elect,new, Durhuus:2009sm}.  In the 2d case the multigraphs have $d_s=2$, so the CDTs have $d_s\le 2$ and it is believed, although not proven, that $d_s=2$.  Here we pursue this intuition and show that it can account for the behaviour of the spectral dimension observed in numerical simulations of 4d CDT in the physical phase where vertices of arbitrarily  high degree are not seen \cite{HLrel}. In essence we argue that  a reduced, or ``toy", model based on an ensemble of multigraphs obtained from radial reduction of the CDTs  carries all the information about spectral dimension; it does not of course carry information about everything else as many degrees of freedom have been integrated out.

%

\section{Multigraphs as toy models for causal quantum gravity}
{The measure $\meas$ of the multigraph ensemble related to four-dimensional CDT is not known analytically; instead we will show that to determine the spectral dimension it is sufficient  to introduce  three assumptions concerning the volume and resistance growth in this measure.  

Computer simulations in four-dimensional CDT \cite{Ambjorn:2004aa} show that for a graph  $T$ with maximum distance from the root $t$ the number of 4-simplices is $\avg{ N_4(t)}_Z = c\, t^4$ so the average number of time-like edges is bounded by 
\bea\label{simulation-vol}
\avg {B(t)}_{\meas} \equiv  \avg{ \sum_{n=0}^{t}  L_n}_{\meas}  \leq c' t^4.
\eea
We will assume that the expectation of the number of time-like edges takes the form
\bea \label{assump-connectivity}
\avg{L_N}_{\meas} 
\simeq \nu  N^{3-\epsilon} +  N
\eea
where ``$\simeq$" denotes equality up to a multiplicative constant and
 $\epsilon>0$ can be taken arbitrarily small. (The use of $\epsilon$ here is for purely technical reasons and for all practical purposes one can think of it as being zero.) It follows from \eqref{assump-connectivity} that there are positive constants $\underline{c}, \bar{c}$ such that  
\bea \label{assump-ball}
\underline{c}N\avg{L_N}_{\meas} \leq \avg{B(N)}_{\meas} 
\leq  \bar{c}N\avg{L_N}_{\meas} 
\eea
which   is consistent with \eqref{simulation-vol}. The $N^2$ sub-leading term is absent in   \eqref{assump-connectivity} as to survive the continuum limit it would have to couple with $\sqrt{\nu}$, 
which in turn would imply its appearance   in the Euclidean Einstein-Regge action.
Our other assumptions concern the size of the fluctuations in $L_N$. Defining the resistance from a vertex at distance $N$ to infinity to be $R(N) \equiv \sum_{n=N}^{\infty} \frac{1}{L_n}$, we assume
\bea
\label{resistance}
R(N) &\leq & \frac{N}{ \avg{L_N}_{\meas}}  \psi _{+}(\sqrt{\nu} N^{\frac{2-\epsilon}{2}})\\
\label{connectivity}
L_N &\leq& \avg{L_N}_{\meas} \psi (\sqrt{\nu} N^{\frac{2-\epsilon}{2}})
\eea
at large $N$ for almost all graphs of the ensemble. In other words, the above bounds are obeyed by any typical configuration, such as those obtained from self-averaging in the computer simulations. 
Upward fluctuations in $L_N$ are controlled by   \eqref{connectivity} and downward fluctuations by  \eqref{resistance} and $\psi (x)$ and $\psi _{+}(x)$ are functions which diverge and vary slowly at $x=0$ and $x= \infty$.\footnote{A function $\psi(x)$ is slowly varying at $x_0$ if $\lim_{x\to x_0} \psi(\lambda x)/\psi(x) =1$ for any $\lambda>0$.} 
 This is the known behaviour of the fluctuations in 2d CDT, where  $\psi (x) \simeq \psi _{+}(x) \simeq {|\log x|} $ \cite{Durhuus:2009sm}, and consistent with the simulation results in 4d.
Note that the finite resistance $R(N)$ implies that this multigraph ensemble is non-recurrent, i.e. $d_s\geq 2$.

\section{
Spectral dimension and  dimensional reduction}
We now show that the multigraph model whose measure $\meas$ has the properties \eqref{assump-connectivity}, \eqref{resistance} and \eqref{connectivity}  has a scale dependent spectral dimension which is two at small scales and four at large scales, as observed in the numerical simulations of 4d CDT. In order to do this we use the formalism developed in  \cite{Atkin:2011ak}  to understand scale dependent spectral dimension on graphs. Full technical details, such as mathematically rigorous proofs, may be found in \cite{new}.}

Let $p_M(t)$ be the probability  that a random walker on a fixed multigraph $M$ returns to the root $0$ after time $t$. The spectral dimension is defined through $p_M(t)$ via the relation $p_M(t) \sim t^{-\dS/2}$ at large time. More precisely, using the generating function for the return probabilities
\bea
\label{definition of Q}
Q_M (x) = \sum _{t=0}^{\infty} p_M(t) (1-x)^{t/2}
\eea 
we define the spectral dimension using
\bea
Q_M(x) \sim x^{-1 + \dS/2} \qquad \text{as} \qquad  x\rar 0
\eea
in the case where the random walk is recurrent ($\dS < 2$) and $Q_M(x)$ diverges as $x\to 0$. If the random walk is non-recurrent, $Q_M(x)$ is finite and we define the spectral dimension through the derivative of $Q_M(x)$ of lowest degree which is diverging via the relation
\bea
Q_M^{(k)}(x) \sim x^{-1-k + \dS/2} \qquad \text{as} \qquad  x\rar 0
\eea
for $2k\leq \dS < 2(k+1) $. As we will see later for $\meas$ the expectation value of $Q_M (x)$ is finite while its first derivative diverges. 

Given a fixed multigraph $M$ the probability for a random walker at $n$ to step next to $n+1$ is given by $p_n(M) = L_n/(L_{n-1} + L_n)$ and the probability that the next step is to $n-1$ is $1-p_n(M)$ (note that the probability to move from the root to vertex one is $1$). Then we decompose the random walk into two pieces; a step from vertex  $n$ to  $n+1$, then a random walk returning to $n+1$ and a final step from $n+1$ to n at time $t$. This decomposition relates $Q_{M_n}(x)$ and $Q_{M_{n+1}}(x)$ and the generating function satisfies the following recursion relation \cite{Durhuus:2009sm}
\bea
\label{recursion relation of Q}
\eta _{M_n}(x) =\eta _{M_{n+1}}(x) + \frac{1}{L_n} -x L_{n} \eta _{M_n}(x)\eta _{M_{n+1}}(x) 
\eea 
where $\eta _{M_n}(x)\!\equiv\!  Q_{M_{n}}(x)/L_n$ and ${M_{n}}$ is the multigraph obtained from $M$ by removing the first $n$ vertices and all edges attached to them and relabelling the remaining multigraph. Recall that $Q_M(x)\equiv Q_{M_{0}}(x)$.  
Differentiating \eqref{recursion relation of Q} and iterating we get
\bea
 \abs{\eta '_{M_{0}}(x)} &=& \abs{\eta '_{M_{N}}(x)} \cdot \prod _{n=0}^{N-1} \frac{1 - x L_n \eta _{M_{n}}(x)}{1+x L_n \eta _{M_{n+1}}(x)} \nn \\
&&+\sum _{n=0}^{N-1} \frac{L_n \eta _{M_{n}}(x) \eta _{M_{n+1}}(x)}{1+x L_n \eta _{M_{n+1}}(x)}    \cdot   \prod _{k=0}^{n-1} \frac{1 - x  L_k \eta _{M_{k}}(x)}{1+x L_k \eta _{M_{k+1}}(x)}
\label{iteration of Q'}
\eea
which is the starting point of our proofs.
%
%
%

On the one hand, \eqref{iteration of Q'} can be rearranged using \eqref{recursion relation of Q} and bounded from below  by
\bea \label{etaprime:lower} 
\abs{\eta_0'(x)}&>&\abs{\eta_{N}'(x)} (1-x)^N e^{-2x\sum_{k=0}^{N-1} L_k\eta_{k+1}(x)}+\nn\\
                                  &+&\sum_{n=0}^{N-1} (L_n\eta_{n+1}(x)^2+\eta_{n+1}(x) )(1-x)^n e^{-2x\sum_{k=0}^{n} L_k\eta_{k+1}(x)}. 
\eea
Using the Cauchy--Schwartz inequality on the second term and  \eqref{resistance}--\eqref{connectivity}, one obtains that \eqref{etaprime:lower} is bounded further by
\bea
 \abs{\eta '_0(x)} >  c \frac{\left (1-x N^{*}(x) \psi \left ( \sqrt{\nu} (N^{-})^{1-\epsilon/2}\right ) \psi_+\left (\sqrt{\nu} (N^{-})^{1-\epsilon/2}\right)  \right )^2}{x^2 B(N^*)}
\eea
where $N^* = \lceil bx^{-\half} \rceil> N^{-}$  and $b$ is a positive constant. Note that the second term in the numerator is sub-leading as $x\to0$, since the functions $\psi(x)$, $\psi_+(x)$ are slowly varying. Averaging over the ensemble $\mu(\nu)$,  applying Jensen's inequality and using \eqref{assump-connectivity}--\eqref{assump-ball} we get
\bea \label{Qprime_lower}
\avg{\abs{Q_M'(x)}}_{\meas} > c_{-}  \frac{1}{x^{\frac{\epsilon}{2}}\nu +x }.
\eea

On the other hand, \eqref{iteration of Q'} can be rearranged so that it is bounded from above by
\bea \label{etaprime:upper}
\abs{\eta_0'(x)} &<&\abs{\eta_{N}'(x)} (1-x)^{-N}e^{-2x\sum_{k=0}^{N-1} L_k\eta_{k}(x)} +\nn\\
                           &+&\sum_{n=0}^{N-1}( L_n\eta_{n+1 }(x)^2+\eta_{n+1}(x)) (1-x)^{-n-2}e^{-2x\sum_{k=0}^{n-1} L_k\eta_{k}(x)}.
\eea
Using the fact that $\eta_N(x)$ is a finite convex decreasing function in $x=[0,1)$, i.e. $\abs{\eta_{N}'(x) } < \eta_{N}(0)/x$ and note that the exponentials are bounded by constants for the choice $N=N^*$, one gets 
\bea
\abs{\eta '_{M_0}(x)} <
c'\left ( \frac{  \eta_{M_{N^*}}(0)  }{x} +\sum _{n=0}^{N^*-1} L_n \eta _{M_n}(0)\eta _{M_{n+1}}(0) \right )
\eea
where $c$, $c'$ are positive constants. Noting that $\eta _{M_N} (0) = R(N)$ is the (finite) resistance, taking the expectation value and using  the fact that 
\bea
 \left \langle\sum _{n=0}^{N^*-1} L_n \eta _{M_n}(0)\eta _{M_{n+1}}(0) \right \rangle_{\mu(\nu)}< \textrm{const} + c_3 \frac{{N^*}^3}{\avg{L_{N^*}}_{\mu}} \psi ^2_{+}(\sqrt{\nu}{N^*}^{1-\epsilon/2})
 \eea
  together with \eqref{assump-connectivity}--\eqref{connectivity} we finally get
\bea \label{Qprime_upper}
\avg{\abs{Q_M'(x)}}_{\meas} < c_{+} \frac{\psi _{+}^2\left (\nu^{\frac{1}{2}} x^{-\frac{1}{2}+\frac{\epsilon}{4}} \right )}{x^{\frac{\epsilon}{2}}\nu + x}.  
\eea
Combining \eqref{Qprime_lower} and \eqref{Qprime_upper} in a compact form, we write
\bea \label{Qprime_compact}
\avg{\abs{Q_M'(x)}}_{\meas} \sim \frac{1}{\nu x^{\frac{\epsilon}{2}}+ x}.
\eea

We obtain the scaling limit by taking the lattice spacing $a\to0$ and following the prescription of \cite{Atkin:2011ak} define 
\bea
\label{definition_Qprime}
\abs{ \tilde Q'(\xi, G) }\equiv \lim _{a \rar 0 } \left(\frac{a}{G}\right) \avg{\abs{Q_M'(x=a \xi)}}_{\mu (\nu)} 
\eea
with $\nu=a^{1-\frac{\epsilon}{2}}/G$. Using  \eqref{Qprime_lower} and \eqref{Qprime_upper} we now have that
\bea
\label{result__Qprime}
c_{-} \frac{1}{\xi^{\frac{\epsilon}{2}}+G \xi}\leq\abs{ \tilde Q'(\xi, G)}\leq c_{+} \frac{\psi _{+}^2(G^{-\frac{1}{2}} \xi^{-\frac{1}{2}+\frac{\epsilon}{4} })}{\xi^{\frac{\epsilon}{2}}+  G \xi}.
\eea
Eq. \eqref{result__Qprime} implies that $d_s=2$ in the short walk limit (i.e. $\xi\to\infty$) and  $d_s=4-\epsilon$ in the long walk limit  (i.e. $\xi\to0$) for $\epsilon$ arbitrarily small (see \cite{ Atkin:2011ak} for the detailed correspondence between walk length and $\xi$). At this point we should mention that it is a highly non-trivial fact that there exists a non-trivial limit \eqref{definition_Qprime}. The first example of the existence of such a limit (in the recurrent case) was given in the context of random combs \cite{Atkin:2011ak}.

We can extract from \eqref{Qprime_compact} the average return probability as a function of large walk length. 
%
In particular, from \eqref{definition of Q} we get
\bea \label{1-x_times_Qprime}
(1-x)\avg{\abs{Q_M'(x)}}_{\meas} = \sum _{t=0}^{\infty} \frac{t}{2} \avg{p_M(t)}_{\meas}  (1-x)^{t/2}.
\eea
From \eqref{Qprime_compact} the left hand side of \eqref{1-x_times_Qprime} reads
\bea
(1-x)\avg{\abs{Q_M'(x)}}_{\meas} \sim L \left ( \frac{1}{1-\sqrt{1-x}}\right ) \, \, x^{-\epsilon/2} \qquad \text{as} \qquad x \to 0,
\eea 
where $L \left ( \frac{1}{1-\sqrt{1-x}}\right ) = \frac{1-x}{\nu + x^{1-\epsilon/2}}$ and $L(y)$ is a slowly varying function at infinity. Using a Tauberian theorem (chapter XIII, \cite{Feller:1971}) one gets (setting $\epsilon$ to zero in the expressions below to simplify the discussion)
\bea      \label{pav}
\avg{p_M(t)}_{\meas}  \sim 
\frac{2}{ t^2}  \left( \frac{\nu+1}{(1-1/t)^{2}} -1\right)^{-1} 
\eea
as $t \rar \infty $. Scaling $t(a) = \lfloor \sigma/a \rfloor$ and $\nu(a) = a/G$ as before one obtains the probability density of the continuous diffusion time $\sigma$ through
\bea \label{pav-scale}
\tilde{P}(\sigma) \equiv\lim _{a \rar 0} \left ( \frac{a}{G} \right )^{-1} \avg{p_M(t)}_{\mu(\nu )} \sim   \frac{2 G^2}{\sigma^2}  \frac{1}{ 1 + 2G / \sigma}.
\eea
This is precisely expression \eqref{intro-P} which was conjectured in \cite{Ambjorn:2005db} as the behaviour of the continuum return probability density for diffusion on four-dimensional CDT.  It yields the scale dependent continuum spectral dimension
\bea
D_s(\sigma) = 4\left(1- \frac{1}{2+ \sigma/G} \right)   
\eea
consistent with the numerical results.

\section{Discussion}
The multigraph ensemble which describes radially reduced 4d CDT provides some physical insight into the degrees of freedom which determine  the spectral dimension in the physical phase. Firstly the fact that the  behaviour of the return probability density \eqref{intro-P} implied by simulations \cite{Ambjorn:2005db} can be reproduced strongly suggests that the detailed structure of the spatial hypersurfaces is not important; it is the behaviour of the number of time-like edges $L_n$ which is crucial. 
To determine the spectral dimension on the multigraphs it is sufficient to know the volume growth and the resistance behaviour  reflected in the assumptions \eqref{assump-connectivity}, \eqref{resistance} and \eqref{connectivity}. These are all motivated from robust results in lower-dimensional studies \cite{new, Durhuus:2009sm}  but it is a non-trivial result of this letter  that the continuum limit exists and that one can perform it exactly  to obtain the  return probability density \eqref{pav-scale} and show that there is a scale dependent spectral dimension varying from four at large scales to two at small scales.

These results show that the intuition about random walks on sliced graphs described earlier leads to a consistent picture in which the computer observations of a scale dependent spectral dimension can be related to the relatively simple question of the distribution of time-like edges in the CDT providing evidence they could be a real continuum physical phenomenon rather  than a consequence of finite size effects.  While this study only requires the  averages of functions of $L_n$  it  indicates that  further light may be shed on  the mechanisms of dynamical dimensional reduction in four-dimensional CDT  by investigating the distribution and correlations of the $L_n$ in the numerical simulations. Understanding these distributions would help towards an analytical solution of the full four-dimensional model.

\section*{
Acknowledgements}
%
%
GG acknowledges the support of the A.G. Leventis Foundation and  A.S. Onassis Public Benefit Foundation grant F-ZG 097/ 2010-2011. JFW and SZ are supported by EPSRC grant EP/I01263X/1 and STFC grant ST/G000492/1. SZ would like to acknowledge the support of a Visiting Scholarship at Corpus Christi College, Oxford University.

\end{document}